\documentclass[12pt,preprint]{aastex}
\usepackage{amsmath}
\shorttitle{CAL 87}
\shortauthors{Ablimit \& X.-D Li.}

\begin{document}


\title{ The orbital period evolution of the supersoft X-ray source CAL 87 }
\author{Iminhaji Ablimit\altaffilmark{1} and Xiang-Dong Li\altaffilmark{2,3}}
%\email{$^\dagger$iminhajia@gmail.com}
\altaffiltext{1}{Key Laboratory for Optical Astronomy, National Astronomical Observatories,
Chinese Academy of Sciences, Beijing 100012, China}
\altaffiltext{2}{Department of Astronomy, Nanjing University, Nanjing 210046, China}
\altaffiltext{3}{Key Laboratory of of Modern Astronomy and Astrophysics,
Ministry of Education, Nanjing 210046, China}

%\date{}
%\pagerange{\pageref{firstpage}--\pageref{lastpage}} \pubyear{2007}
%\maketitle
%\label{firstpage}

\begin{abstract}
CAL 87 is one of the best known supersoft X-ray sources.
However, the measured masses, orbital period and orbital period
evolution of CAL 87 cannot be addressed by the standard
thermal-timescale mass-transfer model for supersoft X-ray sources.
In this work we explore the orbital evolution of CAL 87 with both
analytic and numerical methods. We demonstrate that the
characteristics mentioned above can be naturally accounted for by
the excited-wind-driven mass-transfer model.

\end{abstract}


\section{Introduction}
Supersoft X-ray sources (SSSs) are characterized by black body-like
spectra with temperatures $\sim 20-100$ eV and  X-ray luminosities
$\sim 10^{35}-10^{38}\, {\rm erg\,s^{-1}}$\citep{gr1996}. \citet
{va92} suggested that the supersoft X-ray emission stems from stable
nuclear burning on the surface of white dwarfs (WDs), which requires
a high mass transfer ($\gtrsim 10^{-7}\,M_\sun\,{\rm yr}^{-1}$) from
the donor star in an interacting binary.  This leading scenario then
involves mass transfer, which is unstable on a thermal timescale,
from a more massive main-sequence (MS) or subgiant donor star
\citep[see][for a review]{Kah1997}.

CAL 87 is a well known SSS in the Large Magellanic Cloud
\citep{C90,hu1998} with X-ray luminosity $\sim 4\times10^{36} {\rm
erg\,s^{-1}}$ \citep{S04}. It is a close binary with orbital period
of $P_{\rm orb} = 10.6$ hr \citep{A97}. The WD is very massive with
mass $\sim 1.35M_\sun$ \citep{S04}, but the donor mass ($\sim 0.34
M_\sun$) is quite low \citep{Oliveira2007}. Moreover,
\citet{Oliveira2007} showed that the orbital period of CAL 87 was
increasing at a rate of $\dot{P}_{\rm orb} =
7.2(\pm1.3)\times10^{-10}$ ss$^{-1}$. A more recent analysis by
\citet{r14} refined the value of $\dot{P}_{\rm orb}$ to be
$6(\pm2)\times10^{-10}$  ss$^{-1}$. CAL 87 is thought to be the
prototype of SSSs which predicts a decreasing orbital period,
but both the small mass ratio and the expanding
orbit do not fit the standard picture of SSSs.

\citet{van1998} and \citet{King1998} have alternatively proposed a
self-excited wind model for SSSs. In this model, the companion star
is irradiated by the soft X-rays from the SSS, exciting strong winds
that drive Roche-lobe overflow (RLOF) at a high rate to sustain
steady hydrogen burning on the accreting WD, even when the companion
star is less massive than the WD. In the binaries with $M_2/M_1<1$,
the wind-driven process is expected to be triggered by: (1) a long phase of residual
hydrogen burning after a mild shell flash, (2) a late helium shell flash
of the cooling white dwarf, after the system has already come into contact
as a cataclysmic variable \citep{King1998}.
In this work, we will examine
whether this wind-driven mass transfer model can explain the orbital
evolution of CAL 87. We present an analytical derivation
of the mass-transfer rate in section 2
and numerical calculations of this rate in section 3.
The results are summarized in section 4.

\section{Analytical Method}
\label{sec:model} The total orbital angular momentum of a binary is
give by
\begin{equation}
J =\mu ({\rm G} M a)^{1/2}={\rm G}^{2/3}{M}^{2/3}(\frac{P_{\rm
orb}}{2\pi})^{1/3},
\end{equation}
where $\mu=M_1M_2/M$ is the reduced mass, $M=M_1+M_2$ is the total
mass ($M_{\rm 1}$ and $M_{\rm 2}$ are the masses of the WD the donor
star, respectively), and $a$ is the binary separation. From Eq.~(1)
the rate of change of the orbital angular momentum can be
derived to be,
\begin{equation}
\frac{\dot{J}}{J} = \frac{\dot{M}_{\rm 1}}{M_{\rm 1}} +
\frac{\dot{M}_{\rm 2}}{M_{\rm 2}} - \frac{1}{3}\frac{\dot{M}}{M} +
\frac{1}{3}\frac{\dot{P}_{\rm orb}}{P_{\rm orb}}.
\end{equation}
Mass loss from the donor star is composed by two parts,
\begin{equation}
{\dot{M}_{\rm 2}}={\dot{M}_{\rm 2tr}} + {\dot{M}_{\rm 2w}},
\end{equation}
where $\dot{M}_{\rm 2tr}$ is the mass transfer rate through RLOF,
and $\dot{M}_{\rm 2w}$ is the excited wind loss rate from the donor
star. Only part of the transferred material from the donor star can
be used to increase the WD mass. The mass growth rate of the WD is,
\begin{equation}
{\dot{M}_{\rm 1}} = \alpha_{\rm H} \alpha_{\rm He} |{\dot{M}_{\rm
2tr}}|,
\end{equation}
where $\alpha_{\rm H}$ and $\alpha_{\rm He}$  are the accumulation
ratios for hydrogen and helium burning, respectively. The excess
material is assumed to be ejected from the surface of the WD. The
total mass loss rate is then,
\begin{equation}
{\dot{M}} = (1 - \alpha_{\rm H} \alpha_{\rm He}) {\dot{M}_{\rm 2tr}}
+ {\dot{M}_{\rm 2w}}.
\end{equation}
Thus the rate of angular momentum loss caused by the mass loss from
the WD and the donor star is
\begin{equation}
\dot{J}_{\rm ML} = (1 - \alpha_{\rm H} \alpha_{\rm He})
{\dot{M}_{\rm 2tr}} a_1^2 \omega + {\dot{M}_{\rm 2w}}a_2^2 \omega
\end{equation}
where $a_1 =(\frac{M_2}{M})a$ and $a_2 =(\frac{M_1}{M}) a$ are the
distances between the WD and the center of mass, and between the
donor star and the center of mass, respectively, $\omega=2\pi/P_{\rm
orb}$ is the orbital angular velocity.

For angular momentum loss caused by magnetic braking (MB) we adopt
the saturated law proposed by \citet{A03} and \citet{s00}
\begin{equation}
\dot{J}_{\rm MB}  =\left\{
\begin{aligned}[l]
-K \omega^3 \left(\frac{R_2}{R_{\sun}}\right)^{1/2}
\left(\frac{M_2}{M_{\sun}}\right)^{-1/2},  \ {\rm if}\ \omega \leq \omega_{\rm{crit}}, \\
-K \omega_{\rm{crit}}^2 \omega
\left(\frac{R_2}{R_{\sun}}\right)^{1/2}
\left(\frac{M_2}{M_{\sun}}\right)^{-1/2},  \ {\rm if}\ \omega >
\omega_{\rm{crit}},
\end{aligned}
\right.
\end{equation}
where $K=2.7\times10^{47}$ gcm$^2$ s, and $\omega_{\rm{crit}}$ is
the critical angular velocity at which the angular momentum loss
rate reaches a saturated state.

Combining Eqs. (2)-(7) we get for the rate of change of the orbital
period \textbf{:}
\begin{equation}
\frac{\dot{P}_{\rm orb}}{P_{\rm orb}} = - \frac{[3{M_1}^2 +2(1 -
\alpha_{\rm H} \alpha_{\rm He})M_1 M_2 - 3{M_2}^2] {\dot{M}_{\rm
2tr}} + 2M_1 M_2{\dot{M}_{\rm 2w}}}{M_1 M_2 M}+3\frac{\dot{J}_{\rm
MB}}{J}.
%3K\omega_{\rm c} (\frac{R_2}{R_\odot})^{0.5} (\frac{M_2}{M_\odot})^{-0.5}
%(\frac{4{\pi}^2}{G})^{2/3} \frac{M^{1/3}}{M_1 M_2} \times {P_{\rm orb}}^{-4/3}
\end{equation}
As an order of magnitude estimate, if we take the parameters of CAL
87 (i.e., $M_{\rm 1}=1.35 M_\sun$, $M_{\rm 2}=0.34 M_\sun$,  and
$P_{\rm orb} = 10.6$ hr) and assume ${\dot{M}_{\rm 2w}}\simeq
{\dot{M}_{\rm 2tr}}\simeq -1\times 10^{-7}M_\sun\,{\rm yr}^{-1}$,
$\alpha_{\rm H} \alpha_{\rm He}\sim 0.5$, then we obtain ${\dot{P}_{\rm
orb}}\sim {9.98\times 10^{-10}}$ ss$^{-1}$, roughly consistent with
the observed value.
%This rate just fits the observational rate of CAL 87.

Figure 1 shows how for the above-adopted masses \&
orbital period ${\dot{P}_{\rm orb}}$ depends on the mass-transfer
rate (left panel) and for $M_{\rm 1}=1.35 M_\sun$
how it depends on the mass ratio $q=M_2/M_1$ (right
panel) with $\alpha_{\rm H} \alpha_{\rm He}=0.5$. In the left panel
it is seen that the orbital period increases when $\dot{M}_{\rm
tr}>2\times 10^{-8}\,M_{\sun}$yr$^{-1}$, and ${\dot{P}_{\rm orb}}$
increases with increasing mass-transfer rate. In the right panel
we adopt $M_{\rm 1}=1.35 M_\sun$ and ${\dot{M}_{\rm 2w}}=
{\dot{M}_{\rm 2tr}}={- 1\times 10^{-7}M_\sun\,{\rm yr}^{-1}}$.
One observes that the orbital period always
increases when $q<1$, and ${\dot{P}_{\rm
orb}}$ decreases with increasing $q$.

\section{Numerical calculation}

\cite{van1998} and \cite{King1998} suggested that the soft X-ray
radiation from an accreting WD may heat the donor star and produce a
strong stellar wind from the heated side of the donor star. If the
wind carries away the specific angular momentum of the donor star,
mass transfer will be driven at a rate comparable with the wind loss
rate. The relation between the mass transfer rate $\dot{M}_{\rm tr}$
and the wind loss rate $\dot{M}_{\rm w}$ obeys
\begin{equation}
\dot{M}_{\rm w}\simeq (3.5\times 10^{-7}\,M_\sun\,{\rm yr}^{-1})
(\frac{M_2}{M_\sun})^{5/6}(\frac{M}{M_\sun})^{-1/3} (\eta_{\rm
s}\eta_{\rm a})^{1/2}\phi (\frac{\dot{M}_{\rm
tr}}{10^{-7}\,M_\sun\,{\rm yr}^{-1}})^{1/2},
\end{equation}
for $M_2\lesssim M_1$; and
\begin{eqnarray}
\dot{M}_{\rm w}\simeq (3.5\times 10^{-7}\,M_\sun\,{\rm yr}^{-1})
(\frac{M_2}{M_\sun})^{0.95}(\frac{M}{M_\sun})^{-1/3}
(\frac{M_1}{M_\sun})^{-0.12}(\eta_{\rm s}\eta_{\rm a})^{1/2}\phi
(\frac{\dot{M}_{\rm tr}}{10^{-7}\,M_\sun\,{\rm yr}^{-1}})^{1/2},
\end{eqnarray}
for $M_2\gtrsim M_1$. Here $\eta_{\rm s}$ is the efficiency of the
WD's spectrum in producing ionizing photons normalized to the case
of supersoft X-ray temperatures of about $10^5$ K, $\eta_{\rm a}$ measures
the luminosity per gram of matter accreted relative to the value for
H shell burning, and $\phi$ parameterizes the fraction of the
donor's irradiated face and the fraction of the wind mass escaping
the system.

Taking into account the occurrence of irradiation-excited winds and
their influence on the binary evolution, we investigate the mass
transfer processes of a binary consisting of a CO WD and a MS
companion star with an updated version of Eggleton's stellar
evolution code \citep{Eggleton1971,Eggleton1973}. In addition,
angular momentum loss caused by gravitational wave radiation
\citep{ll75} and MB \citep{A03,s00} is also included in the
calculation. The growth of the WD mass depends on the
accumulation efficiencies of hydrogen- and helium-rich matter.
%\begin{equation}
%\dot{M}_{\rm WD}=\eta_{\rm H}\eta_{\rm He}\dot{M}_{\rm tr},
%\end{equation}
Here we adopt the results of \citet{Pralnik1995} and
\citet{Yaron2005} for the $\alpha_{\rm H}$, and the prescriptions in
\citet{Ka2004} for $\alpha_{\rm He}$.

We have performed numerical calculations of the evolution for a grid
of binaries. To illustrate the possible formation history of CAL 87,
we consider a binary consisting of a CO WD of initial mass $1.16
M_\odot$ and  a MS companion star of initial mass $1.1\, M_\odot$
with an initial orbital period 0.3 day. The detailed calculated
results are as follows.

%\subsection{The initial parameter space for AIC}
Figure 2 shows the RLOF mass-transfer rate (the black line) and the
wind loss rate (the red line) as a function of time. Although less
massive than the WD, the companion star is able to transfer mass to
the WD material at a high enough rate ($\gtrsim10^{-7} M_\sun\,{\rm
yr}^{-1}$), with the help of the excited wind. In Figure 3, we plot
the evolution of the donor's mass (black dashed line) and the WD's
mass (red dotted line). The WD can finally grow in mass up to the
Chandrasekhar limit mass and explode as a supernova Ia. Figure 4
shows the evolution of the orbital period (left) and its
derivative (right). Initially MB-induced angular momentum loss
causes the orbit to shrink. When the excited wind loss is
sufficiently strong, it starts to dominate the orbital evolution.
Rapid mass transfer and mass loss cause the orbit to expand with
time. At the time $t\sim 4\times 10^6$ yr (or $\log t\, ({\rm yr})
\sim 6.6$),
both the orbital period and its derivative are in agreement with
observed values indicated by red dashed lines.

\section{Concluding remarks}

\citet{r14} derived a rate $\dot{P}_{\rm orb} =
6(\pm2)\times10^{-10}$ for the change in the orbital period of CAL
87 from its eclipse maps. In this paper, we employ both analytical
and numerical methods to show that the excited wind-driven mass
transfer model \citep{van1998,King1998} is able to reproduce
observed characteristics of CAL 87. Other evidence supporting this
model comes from the fact that there is no significant variation of
the emission lines of CAL 87 even during eclipses, suggesting that
the lines are formed in an extended region a circumbinary corona,
which could be fed by winds from both the disk and the donor star
\citep{r14}.

\citet{r14} also mentioned that CAL 87 actually displayed cyclic
orbital period changes, with $\dot{P}_{\rm orb}$ changing from
positive to negative value of $-2.2(\pm0.2)\times10^{-10}$, and they
speculated that the latter may be induced by MB coupled with strong
winds. This would require a very strong magnetic field for the donor
star: the negative value of the observed $\dot{P}_{\rm orb}$ can be
accounted for if we amplify the traditional MB-induced $\dot{J}$ by
a factor of $\sim 240$. An alternative explanation is that mass loss
during the shell burning flashes may decrease the orbital period.
Assuming that during the flashes a fraction $\delta$ of the accreted
matter escapes from the binary system through the outer Lagrangian
($L_3$) point, then the angular momentum loss rate is,
\begin{equation}
\frac{\dot{J}_{L_3}}{J}=-\delta \frac{{a_{L_3}}^2}{a^2}
\frac{M}{M_1} \frac{\dot{M}_{\rm 2tr}}{M_2}
\end{equation}
where $a_{L_3}$ is the distance between the mass center of the
binary and the $L_3$ point. Taking $\delta\sim 0.15-0.45$
\citep{sh12}, we find that when the mass-transfer rate is $\gtrsim
4\times10^{-8}\,M_\sun\,{\rm yr}^{-1}$,
the escaping matter may shrink the orbit at a rate
comparable with observed.

\acknowledgments This work was supported by the Natural Science
Foundation of China under grant number 11133001 and 11333004, and
the Strategic Priority Research Program of CAS under grant No.
XDB09000000.

%\bibliographystyle{apj}
%\bibliography{haji}

\begin{figure}
\centering
\includegraphics[totalheight=2.5in,width=6in]{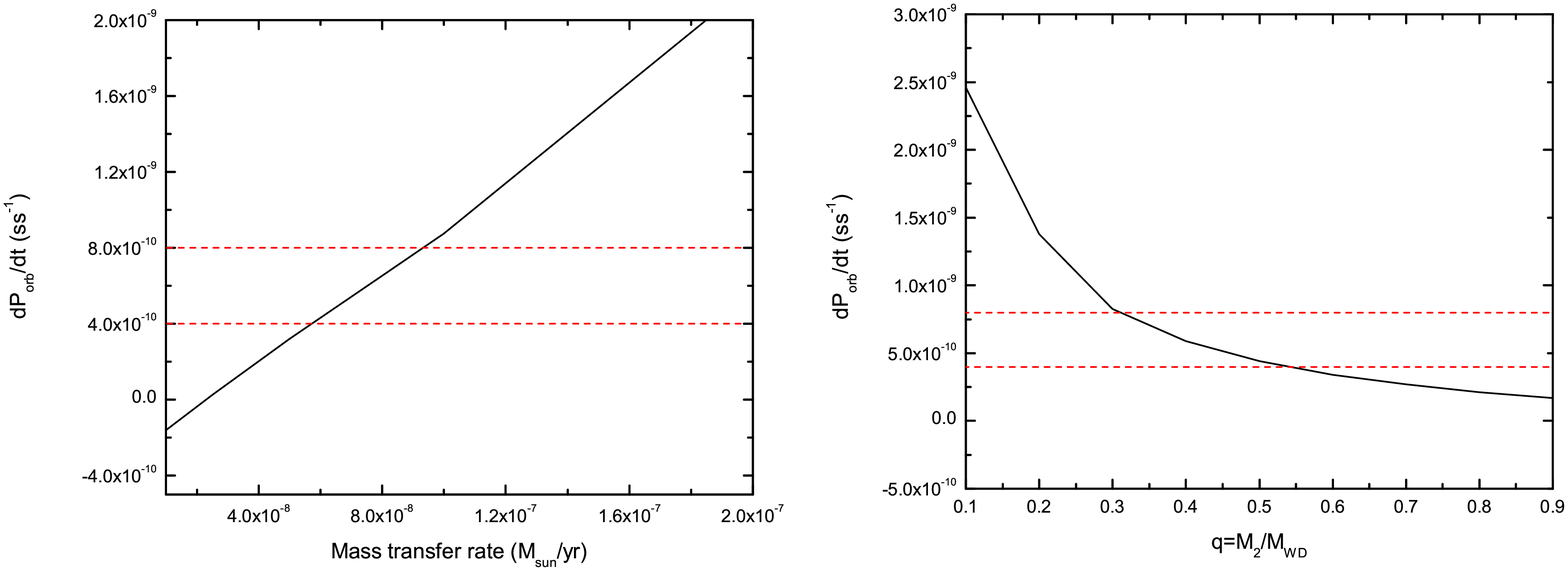}
\caption{Dependance of ${\dot{P}_{\rm orb}}$ on the mass transfer
rate and the mass ratio. The red dashed lines represent the observed
values of CAL 87.}\label{fig:1}
\end{figure}

\clearpage

\begin{figure}
\centering
\includegraphics[totalheight=4.5in,width=5.5in]{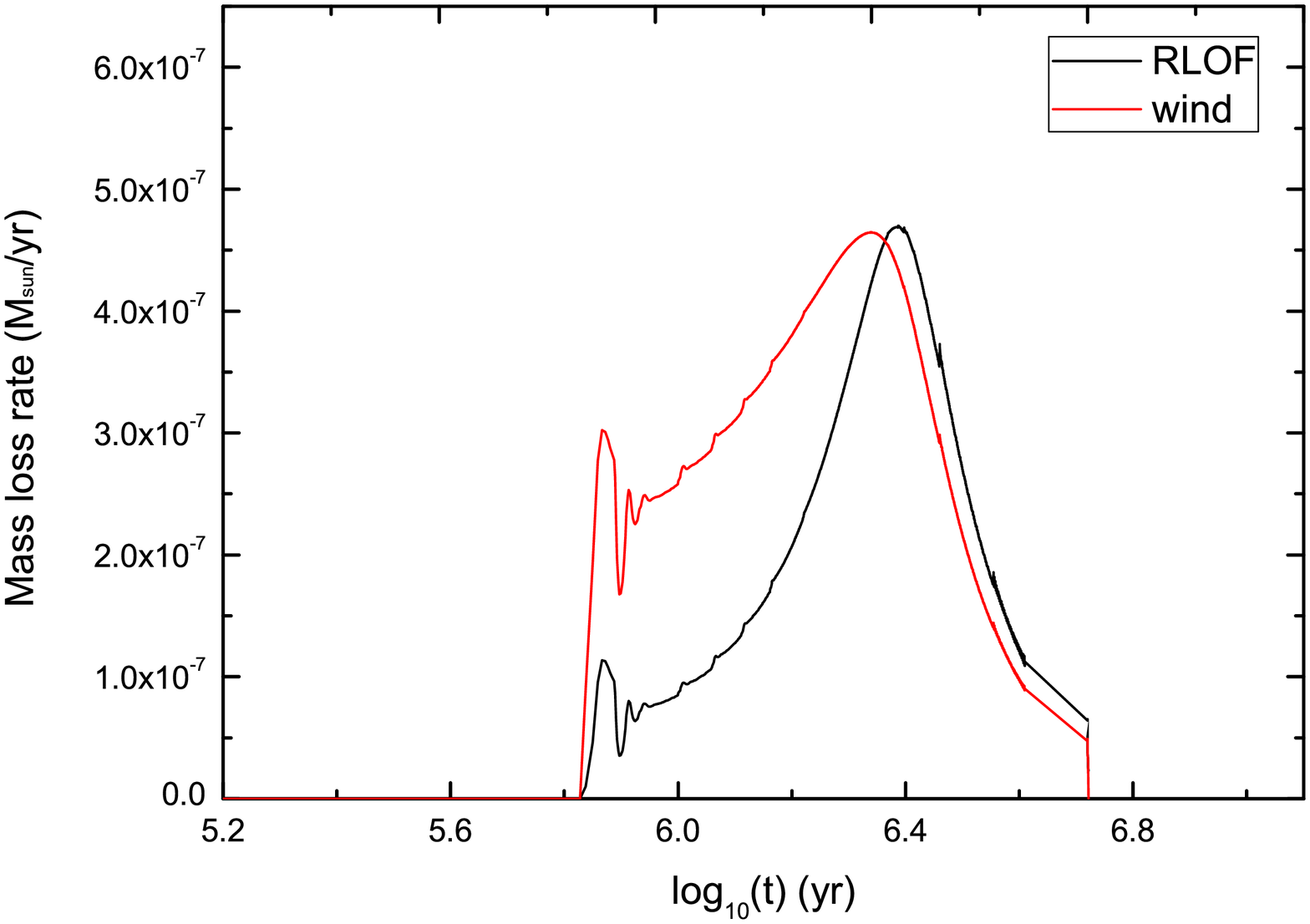}
\caption{Evolution of the RLOF mass-transfer rate (black line) and
the excited wind loss rate (red line).}\label{fig:1}
\end{figure}

\clearpage

\begin{figure}
\centering
\includegraphics[totalheight=4.5in,width=5.5in]{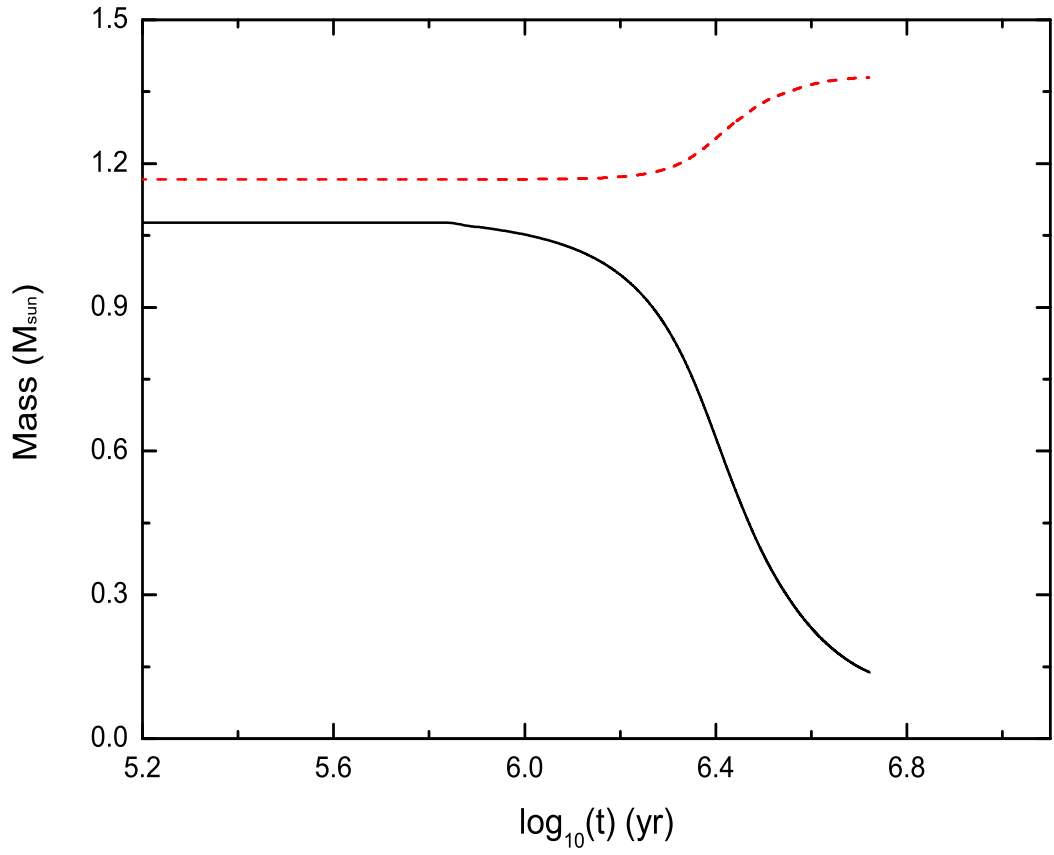}
\caption{Evolution of the donor's mass (black line) and the WD mass
(red line).}\label{fig:1}
\end{figure}

\clearpage

\begin{figure}
\centering
\includegraphics[totalheight=2.5in,width=6in]{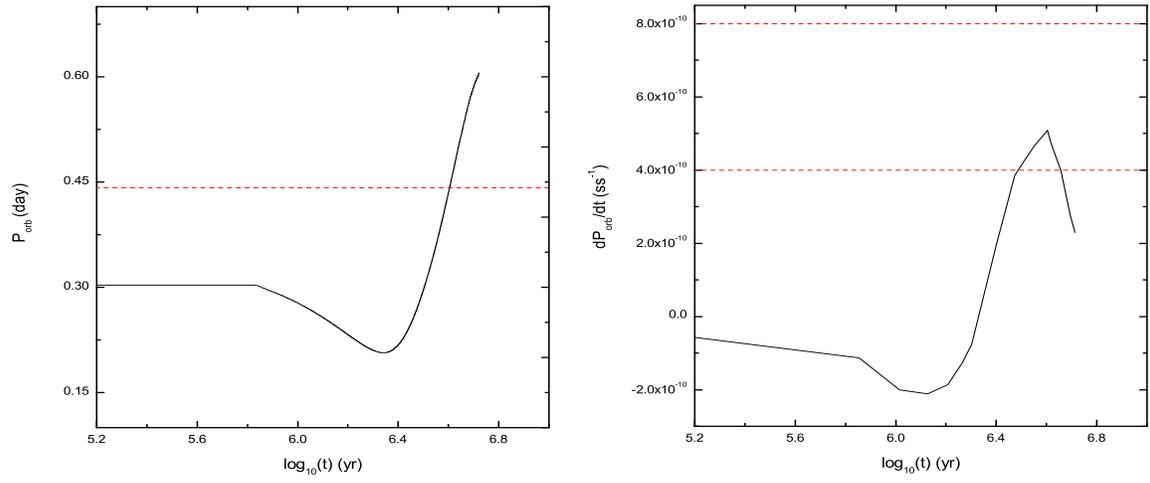}
\caption{Evolution of the orbital period (left) and its derivative
(right).}\label{fig:1}
\end{figure}

\end{document}